\documentclass[12pt,halfline,a4paper]{ouparticle}
\newtheorem{theorem}{Theorem}[section]

\begin{document}

\title{Topological analysis of entropy measure using regression model for terpyridine complex nanosheet}
\author{%
\name{H. M. Nagesh}
\address{Department of Science and Humanities, \\
PES University, Bangalore, India.}
\email{hmnagesh1982@gmail.com.}}

\abstract{A numerical parameter, known as a topological index, is employed to represent the molecular structure of a compound by considering its graph-theoretical properties. In the study of quantitative structure-activity relationships (QSAR) and quantitative structure-property relationships (QSPR), topological indices are used to predict the physicochemical properties of chemical compounds. Graph entropies have evolved as information-theoretic tools to investigate the structural information of a molecular graph. In this research work, we compute the Nirmala index, the first and second inverse Nirmala index for terpyridine complex nanosheet $TCN_{n,m}$ with the help of its M-polynomial. Further, entropy measures based on Nirmala indices are calculated for terpyridine complex nanosheets. We expand this analysis to include visual comparisons, which could be useful in refining the structure for more effective implementation. Furthermore, using graphical and numerical methods, the correlation and comparison between the Nirmala indices and the corresponding entropy measurements are shown. The relationship between the Nirmala indices and associated entropy measurements is then examined using a regression model.
 }

\date{}

\maketitle

\newpage
\section{Introduction}
\label{sec1}
Let $\Upsilon =(V(\Upsilon), E(\Upsilon))$ be an ordered pair of a simple, connected, and undirected graph with non-empty vertex set $V(\Upsilon)$ and edge set $E(\Upsilon)$. The total number of edges incident to a vertex $v \in V(\Upsilon)$ is known as the degree of the vertex $v$ and is denoted as $d_{\Upsilon}(v)$. Let $e=pq$ represent an edge of the graph $\Upsilon$, where $p$ and $q$ are the end vertices of the edge $e$. The study and modeling of the structural characteristics of chemical compounds is the focus of the area of chemical graph theory. Chemical compounds are depicted here as graphs, with atoms serving as nodes and chemical bonds serving as edges between atoms.  Molecular structures are mathematically investigated in this field of study using theoretical, computational, and graphical methods \cite{1}. 

Topological indices are mathematical parameters that are obtained from a chemical compound's molecular graph. These indices are used for predicting the substance's physical attributes, chemical composition, and biological activity. Their importance is particularly evident in the context of quantitative structure-property relationship (QSPR) and quantitative structure-activity relationship (QSAR) investigations \cite{2,3}.

The topological index based on degrees, defined on the edge set $E(\Upsilon)$ of a graph $\Upsilon$ \cite{4} is given by
    \begin{equation*}
I(\Upsilon)= \displaystyle \sum_{uv \in E(\Upsilon)} f(d_{\Upsilon}(u), d_{\Upsilon}(v)),
    \end{equation*}
where $f(x, y)$ is a non-negative and symmetric function that depends on the mathematical formulation of the topological index.

Several topological indices have been reported in the literature and have shown benefits in several fields, including drug development, biology, chemistry, computer science, and physics. Proposed by H. Wiener in 1947, the Wiener index was the first and most studied topological index. One noteworthy application is the prediction of paraffin boiling temperatures \cite{5}. Another well-known degree-based topological statistic is the connectedness index, sometimes known as the Randi\'c index and first introduced by Milan Randi\'c in 1975. Its significance for drug development is widely acknowledged \cite{6}. For additional information on topological indices and their applications, readers are referred to \cite{7,8}. 

Numerous efforts have been undertaken to improve the category of degree-based topological indices by the addition of new indices. Kulli in \cite{9} introduced a novel degree-based topological index of a molecular graph $\Upsilon$, called the \emph{Nirmala index} as follows:

\begin{equation}
N(\Upsilon)= \displaystyle \sum_{uv \in E(\Upsilon)} \sqrt{d_{\Upsilon}(u)+d_{\Upsilon}(v)}
    \end{equation}
  \newpage  
Later in 2021, Kulli \cite{10} introduced the notion of the first inverse Nirmala index $IN_{1}(\Upsilon)$ and second inverse Nirmala index $IN_{2}(\Upsilon)$ of a molecular graph $\Upsilon$ as follows.
\begin{equation}
IN_{1}(\Upsilon)= \displaystyle \sum_{uv \in E(\Upsilon)} \sqrt{\frac{1}{d_{\Upsilon}(u)}+\frac{1}{d_{\Upsilon}(v)}}=\displaystyle \sum_{uv \in E(\Upsilon)} \left( \frac{1}{d_{\Upsilon}(u)}+\frac{1}{d_{\Upsilon}(v)} \right)^{\frac{1}{2}}
    \end{equation}
 \begin{equation}
IN_{2}(\Upsilon)= \displaystyle \sum_{uv \in E(\Upsilon)} \frac{1}{\sqrt{\frac{1}{d_{\Upsilon}(u)}+\frac{1}{d_{\Upsilon}(v)}}}=\displaystyle \sum_{uv \in E(\Upsilon)} \left( \frac{1}{d_{\Upsilon}(u)}+\frac{1}{d_{\Upsilon}(v)} \right)^{-\frac{1}{2}}
    \end{equation} 
    
 \vspace{5mm}   
In the past, several topological indices were computed utilizing their conventional mathematical formulation. There are several attempts to look into a compact method that can recover many topological indices of a particular class. In this regard, the concept of a general polynomial was developed, whose values of the necessary topological indices at a given point are produced by its derivatives, integrals, or a combination of both. For instance, the distance-based topological indices are recovered by the Hosoya polynomial \cite{11}, whereas the NM-polynomial generates the neighborhood degree sum-based topological indices \cite{12}. 

The M-polynomial was introduced by Deutsch and Kla\v{z}ar in \cite{13} to find the degree-based topological indices.  \\\\
\textbf{Definition 1.1} (\cite{13}) The M-polynomial of a graph $\Upsilon$ is defined as:
\begin{center}
    $M(\Upsilon; x,y)=\displaystyle \sum_{\delta \leq i \leq j \leq \Delta} m_{i,j}(\Upsilon)x^{i}y^{j}$,
\end{center}
where $\delta = min\{d_{\Upsilon}(u) | u \in V(\Upsilon)\}$, $\Delta = max\{d_{\Upsilon}(u) | u \in V(\Upsilon)\}$, and $m_{ij}$ is the number of edges $uv \in E(\Upsilon)$ such that $d_{\Upsilon}(u)=i, d_{\Upsilon}(v)=j \, (i,j \geq 1)$.

The M-polynomial-based derivation formulae to compute the different Nirmala indices are listed in Table 1.
\newpage
\begin{center}
 \textbf{Table 1}. Relationship between the M-polynomial and Nirmala indices for a graph $\Upsilon$.      
\end{center}

\begin{table}[h!]
\centering
\renewcommand{\arraystretch}{2.5}
\begin{tabular}{||c| c |c| c ||} 
 \hline
 Sl. No & Topological Index & $f(x,y)$ & Derivation from $M(\Upsilon;x,y)$  
 \\ [0.5ex] 
 \hline\hline
 1  & Nirmala index (N) & $\sqrt(x+y)$ & $D_{x}^{\frac{1}{2}}J(M(\Upsilon; x,y))|_{x=1}$ \\
\hline
2  & First inverse Nirmala index $(IN_1)$ & $\sqrt(\frac{x+y}{xy})$ & $D_{x}^{\frac{1}{2}}JS_{y}^{\frac{1}{2}}S_{x}^{\frac{1}{2}}(M(\Upsilon; x,y))|_{x=1}$\\
\hline
 3  & Second inverse Nirmala index $(IN_2)$ & $\sqrt(\frac{xy}{x+y})$ & $S_{x}^{\frac{1}{2}}JD_{y}^{\frac{1}{2}}D_{x}^{\frac{1}{2}}(M(\Upsilon; x,y))|_{x=1}$  \\
 [1ex] 
 \hline
\end{tabular}
\end{table} 
\vspace{5mm}
Here, $D_{x}^{\frac{1}{2}}(h(x,y))=\sqrt{x \cdot \frac{\partial(h(x,y))}{\partial x}} \cdot \sqrt{h(x,y)}$; \\ $D_{y}^{\frac{1}{2}}(h(x,y))=\sqrt{y \cdot \frac{\partial(h(x,y))}{\partial y}} \cdot \sqrt{h(x,y)}$;\\
$S_{x}^{\frac{1}{2}}(h(x,y))=\sqrt{\displaystyle \int_{0}^{x} \frac{h(t,y)}{t} }dt \cdot \sqrt{h(x,y)}$;\\
$S_{y}^{\frac{1}{2}}(h(x,y))=\sqrt{\displaystyle \int_{0}^{y} \frac{h(x,t)}{t} }dt \cdot \sqrt{h(x,y)}$; and $J(h(x,y))=h(x,x)$ are the operators. \\

We refer readers to \cite{14,15,16,17,18,19,20} for additional information on degree-based topological indices utilizing the M-polynomial.

The basic idea of entropy was introduced by Shannon \cite{21}, as a measurement of the unpredictability or uncertainty of the information present in a system, represented by a probability distribution. The structural information of graphs, networks, and chemical structures was then analyzed using entropy. Graph entropies have become more useful in the recent several years in various disciplines, including mathematics, computer science, biology, chemistry, sociology, and ecology. Graph entropy measures can be classified into different types such as intrinsic and extrinsic
measures, and they correspond to the probability distribution with graph invariants (edges, vertices, etc.). For further information on degree-based graph entropy measures and their uses, readers are referred to \cite{22,23,24,25}.

\subsection{Entropy of a graph in terms of edge-weight}
 In 2014, Chen et al. \cite{26} proposed the concept of the entropy of edge-weighted graphs as follows. 
 
 Let $\Upsilon =(V(\Upsilon), E(\Upsilon), \omega(st))$ be an edge-weight graph, where $V(\Upsilon)$ is a set of vertices,  $E(\Upsilon)$ is a set of edges; and $\omega(st)$ denotes the weight of an edge  $st \in E(\Upsilon)$. Then the entropy of a graph in terms of edge weight is defined as, \vspace{3mm} 
\begin{align*}
ENT_{\omega}(\Upsilon)=&-\displaystyle \sum_{s^{'}t^{'} \in E(\Upsilon)} \frac{\omega(s^{'}t^{'})}{\displaystyle \sum_{st \in E(\Upsilon) } \omega(st) } 
log \left(\frac{\omega(s^{'}t^{'})}{\displaystyle \sum_{st \in E(\Upsilon) } \omega(st) } \right) \\
= & -\displaystyle \sum_{s^{'}t^{'} \in E(\Upsilon)} \frac{\omega(s^{'}t^{'})}{\displaystyle \sum_{st \in E(\Upsilon) } \omega(st) } 
\left[log(\omega(s^{'}t^{'})) -log \left(\displaystyle \sum_{st \in E(\Upsilon)} \omega(st) \right) \right] \\
=& log \left(\displaystyle \sum_{st \in E(\Upsilon)} \omega(st) \right) -\displaystyle \sum_{s^{'}t^{'} \in E(\Upsilon)} \frac{\omega(s^{'}t^{'})}{\displaystyle \sum_{st \in E(\Upsilon) } \omega(st) } 
log(\omega(s^{'}t^{'})) 
\end{align*}
Hence,
\begin{align}
ENT_{\omega}(\Upsilon)=& log \left(\displaystyle \sum_{st \in E(\Upsilon)} \omega(st) \right) - \frac{1}{\left(\displaystyle \sum_{st \in E(\Upsilon) } \omega(st) \right)}              \displaystyle \sum_{s^{'}t^{'} \in E(\Upsilon)} \omega(s^{'}t^{'})  log(\omega(s^{'}t^{'}))
\end{align} 

In 2023, Virendra Kumar et al. \cite{27} introduced the notion of the Nirmala indices-based entropy by considering the meaningful information function $\omega$ as a function associated with the definitions of the Nirmala indices as given in equations (1-3).
 \\\\
\textbf{Nirmala entropy}: Let $\omega(st)=\sqrt{d_{\Upsilon}(s)+d_{\Upsilon}(t)}$. Then from the definition of the Nirmala index as given in Equation (1), we have
\begin{center}
$ \displaystyle \sum_{st \in E(\Upsilon)}\omega(st)=\displaystyle \sum_{st \in E(\Upsilon)} \sqrt{d_{\Upsilon}(s)+d_{\Upsilon}(t)}=N(\Upsilon)$  
\end{center}
Hence, using equation (4), the Nirmala entropy of a graph $\Upsilon$ is given by 
\begin{equation}
 ENT_{N}(\Upsilon)=log(N(\Upsilon))  - \frac{1}{N(\Upsilon)}              \displaystyle \sum_{st \in E(\Upsilon)} \sqrt{d_{\Upsilon}(s)+d_{\Upsilon}(t)} \times log (\sqrt{d_{\Upsilon}(s)+d_{\Upsilon}(t)})
\end{equation} \\\\
\textbf{First inverse Nirmala entropy}: Let $\omega(st)=\sqrt{\frac{1}{d_{\Upsilon}(s)}+\frac{1}{d_{\Upsilon}(t)}}$. Then from the definition of the first inverse Nirmala index as given in Equation (2), we have,
\begin{center}
$ \displaystyle \sum_{st \in E(\Upsilon)}\omega(st)=\displaystyle \sum_{st \in E(\Upsilon)} \sqrt{\frac{1}{d_{\Upsilon}(s)}+\frac{1}{d_{\Upsilon}(t)}}=IN_{1}(\Upsilon)$  
\end{center}
Hence, using equation (4), the first inverse Nirmala
entropy of a graph $\Upsilon$ is given by 
\begin{equation}
 ENT_{IN_{1}}(\Upsilon)=log(IN_{1}(\Upsilon))  - \frac{1}{IN_{1}(\Upsilon)}              \displaystyle \sum_{st \in E(\Upsilon)} \sqrt{\frac{1}{d_{\Upsilon}(s)}+\frac{1}{d_{\Upsilon}(t)}} \times log \left( \sqrt{\frac{1}{d_{\Upsilon}(s)}+\frac{1}{d_{\Upsilon}(t)}} \right) 
\end{equation}
\textbf{Second inverse Nirmala entropy}: Let $\omega(st)=\frac{\sqrt{d_{\Upsilon}(s) \cdot  d_{\Upsilon}(t)} }{\sqrt{d_{\Upsilon}(s) + d_{\Upsilon}(t)} }$.
Then from the definition of the second inverse Nirmala index as given in Equation (3), we have,
\begin{center}
$ \displaystyle \sum_{st \in E(\Upsilon)}\omega(st)=\displaystyle \sum_{st \in E(\Upsilon)} \frac{\sqrt{d_{\Upsilon}(s) \cdot  d_{\Upsilon}(t)} }{\sqrt{d_{\Upsilon}(s) + d_{\Upsilon}(t)} }=IN_{2}(\Upsilon)$  
\end{center}
Hence, using equation (4), the second inverse Nirmala
entropy of a graph $\Upsilon$ is given by 
\begin{equation}
 ENT_{IN_{2}}(\Upsilon)=log(IN_{2}(\Upsilon))  - \frac{1}{IN_{2}(\Upsilon)}              \displaystyle \sum_{st \in E(\Upsilon)} \frac{\sqrt{d_{\Upsilon}(s) \cdot  d_{\Upsilon}(t)} }{\sqrt{d_{\Upsilon}(s) + d_{\Upsilon}(t)} } \times log \left( \frac{\sqrt{d_{\Upsilon}(s) \cdot  d_{\Upsilon}(t)} }{\sqrt{d_{\Upsilon}(s) + d_{\Upsilon}(t)} } \right) 
 \end{equation}
 
 \subsection{Methodology}
In this paper, we use graph entropy measures by employing novel information functions derived from various Nirmala indices definitions. We conduct a comprehensive mathematical and computational exploration of these measures within the terpyridine complex nanosheet $TCN_{n,m}$. The rest of the paper is organized as follows: Section 2 begins with foundational preliminaries and discusses the crystallographic structure of the terpyridine complex nanosheet $TCN_{n,m}$. In Section 3, we compute the Nirmala indices using the M-polynomial, which allows us to calculate the Nirmala indices-based entropy measures of the $TCN_{n,m}$. Section 4 compares the Nirmala indices and their associated entropy measures through numerical data and 2D line plots. Section 5 deals with a regression model to illustrate how the estimated Nirmala indices and associated entropy values fit the curve. Lastly, a discussion and a conclusion are presented in sections 6 and 7, respectively.  

\section{Terpyridine Complex Nanosheet Structure}
The molecular structure of a specific terpyridine complex nanosheet from the supra-molecular family is particularly important in electrochemical fields. In supramolecular chemistry, molecular interactions are important since they include a variety of repulsive and attractive interactions such as metal-ligand interactions and hydrogen bonding to form different molecular frameworks and architectures \cite{28}. \newpage 
One of the most widely used combinations of the supramolecular family is the terpyridine complex nanosheet (TCN) \cite{29}. The terpyridine complex is a graphene nanosheet system, which has several fascinating properties such as extensive carrier mobility in the present trend of electrochemical devices \cite{30}. A graphene nanosheet is a novel material with a two-dimensional polymeric structure. These nanosheets are made by chemical or mechanical exfoliation and have a basis in crystalline mother materials. The ability to modify bottom-up nanosheets' shapes and characteristics through component selection and design is by far their greatest advantage over top-down nanosheets. The development of bottom-up nanosheets made of ionic, atomic, and molecular constituents has drawn a lot of interest lately. Bottom-up nanosheets exhibit intriguing features that are not yet fully understood. A recent example of the wide range of nanosheets that are currently being developed is the synthetic dithiolene nickel complex nanosheet, which was produced by utilizing nickel ion as the metal center and benzenehexathiol as an organic ligand. Numerous topological indices about the architectures of graphene nanosheets have been examined in recent chemical graph theory literature \cite{31}. In the latest literature on chemical graph theory, there has been a significant increase in the number of newly introduced topological indices, which enable the classification of physicochemical properties of chemical compounds. 

Renai et al. in \cite{32} investigated and computed topological indices and their corresponding entropy for the terpyridine complex nanosheet $TCN_{n,m}$ which is also known as an electrochromic bottom-up nanosheet with the composition of three-fold symmetric TPY ligands (terpyridyl phenyle and terpyridyl ethynyl with Co and Fe ions) for its future utilization in QSPR and QSAR analysis and its implications in physicochemical and biological aspects. One specific TCN molecule is a crystalline complex made up of a finite number of hydrogen bonds (edges) and atoms (vertices) produced by combining terpyridine TPY with either Fe or Co as a metal-ligand in a three-fold system. From a graphical perspective, the atoms and hydrogen bonds in this context are called $edges$ and $vertices$, respectively. The compound structure $TCN_{n,m}$ has matching regular hexagons with lengths of five $C_5$ and six $C_6$. 

Figure 1 depicts a TCN compound's initial dimensional structure, that is, $TCN_{n,m}$ for $n=m=1$. The $n,m$ dimensional TCN structure $TCN_{n,m}$ is shown in Figure 2. 

\newpage
\begin{figure}[h!]
\centering
\includegraphics[width=175mm]{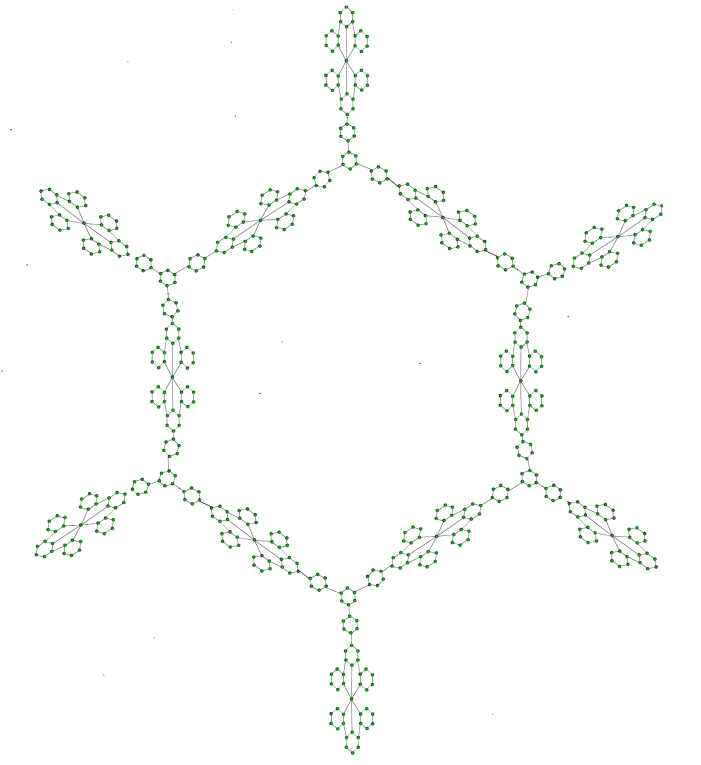}  
  \end{figure} 
  \begin{center}
  Figure 1. Terpyridine complex nanosheet $TCN_{n,m}$ with dimension  $n=m=1$.    
  \end{center}
\newpage
\begin{figure}[h!]
\centering
\includegraphics[width=180mm]{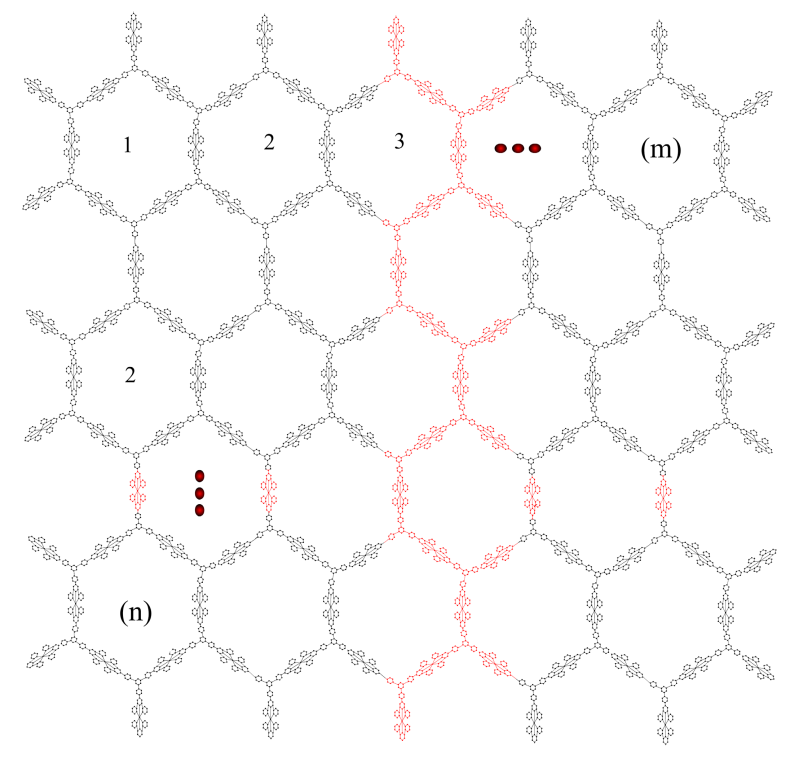}  
  \end{figure} 
  \begin{center}
  Figure 2. $n,m$ dimensional TCN structure $TCN_{n,m}$.    
  \end{center}
  
Motivated by the studies described in \cite{27, 32}, we aim to calculate the Nirmala index, as well as the first and second inverse Nirmala index, for the terpyridine complex nanosheet $TCN_{n,m}$ using M-polynomial. Additionally, entropy measures for terpyridine complex nanosheets are computed using Nirmala indices. 
 \newpage 

\section{Main results}
Unless otherwise mentioned, $TCN_{n,m}$ represents the $n,m$ dimensional terpyridine complex nanosheet in this study. 

In this section, we first find the M-polynomial of terpyridine complex nanosheet $TCN_{n,m}$. Then, we compute the Nirmala index, and first and second inverse Nirmala indices using its M-polynomial.

The size of the $TCN_{n,m}$ molecular graph is given by
\begin{center}
$|E(TCN_{n,m})|=396mn+46n+290m$.    
\end{center} 
The edge set partitions of $TCN_{n,m}$ molecular graph are shown in Table 2.
\vspace{3mm}
\begin{table}[h!]
\centering
\renewcommand{\arraystretch}{2.5}
\begin{tabular}{||c| c |c| c ||} 
 \hline
 Sl. No & Edge set & $(d_{\Upsilon}(r),d_{\Upsilon}(s))$ & Number of repetitions  
 \\ [0.5ex] 
 \hline\hline
 1  & $E_1$  & (2,2) & $96mn+16n+80m$\\
\hline
2  & $E_2$ & (2,3) & $168mn+12n+108m$ \\
\hline
 3  & $E_3$ & (3,3) & $96mn+12n+72m$  \\
 \hline
 4  & $E_4$ & (3,6) & $36mn+6n+30m$  \\
 \hline 
\end{tabular}
\end{table} 
\begin{center}
    \textbf{Table 2}. Edge set partition of $TCN_{n,m}$ molecular graph according to degrees of end vertices of an edge. 
\end{center}

\subsection{Nirmala indices of $TCN_{n,m}$}
We now find the M-polynomial of $TCN_{n,m}$ as follows.
\begin{theorem}
Let $\Upsilon$ be the $n,m$ dimensional terpyridine complex nanosheet $TCN_{n,m} $. Then the M-polynomial of $\Upsilon$ is 
\begin{align*}
M(\Upsilon; x,y)=&(96mn+16n+80m)x^2y^2+(168mn+12n+108m)x^2y^3+ (96mn+12n+72m)x^3y^3\\
&+ (36mn+6n+30m)x^3y^6.
\end{align*}
\end{theorem}
\textbf{Proof.} Let $\Upsilon$ be the $n,m$ dimensional terpyridine complex nanosheet $TCN_{n,m} $.\\ From Table 2,  $\displaystyle \sum_{i=1}^{4}|E_{i}|=396mn+46n+290m$.

Since each vertex of $\Upsilon$ is of degree either $2$ or $3$ or $6$, 
the partitions of edge set $E(\Upsilon)$ are: \\
$E_{1}(\Upsilon):=\{e=ab \in E(\Upsilon): d_{\Upsilon}(a)=2, d_{\Upsilon}(b)=2 \}$;\\
$E_{2}(\Upsilon)):=\{e=ab \in E(\Upsilon): d_{\Upsilon}(a)=2, d_{\Upsilon}(b)=3 \}$;\\
$E_{3}(\Upsilon):=\{e=ab \in E(\Upsilon): d_{\Upsilon}(a)=3, d_{\Upsilon}(b)=3 \}$;\\
$E_{4}(\Upsilon):=\{e=ab \in E(\Upsilon): d_{\Upsilon}(a)=3, d_{\Upsilon}(b)=6 \}$.\\\\
Clearly, $|E_{1}(\Upsilon)|=96mn+16n+80m; |E_{2}(\Upsilon)|=168mn+12n+108m; \\|E_{3}(\Upsilon)|=96mn+12n+72m; |E_{4}(\Upsilon)|=36mn+6n+30m$. Therefore, 
\begin{align*}
M(\Upsilon; x,y)=&\displaystyle \sum_{\delta \leq i \leq j \leq \Delta} m_{i,j}(\Upsilon)x^{i}y^{j} = m_{22}(\Upsilon)x^{2}y^{2}+m_{23} (\Upsilon)x^{2}y^{3}+m_{33}(\Upsilon)x^{3}y^{6} + m_{36}(\Upsilon)x^{3}y^{6} \\ 
 = &(96mn+16n+80m)x^2y^2+(168mn+12n+108m)x^2y^3+ (96mn+12n+72m)x^3y^3\\
   & + (36mn+6n+30m)x^3y^6.
 \end{align*}
  Now we evaluate the Nirmala indices of $TCN_{n,m}$ with the help of its M-polynomial.
\begin{theorem}
Let $\Upsilon$ be the $n,m$ dimensional terpyridine complex nanosheet $TCN_{n,m} $. Then the Nirmala indices of $\Upsilon$ are:
\begin{align*}
a. N(\Upsilon)=&300mn+250m+50n+\sqrt{5}\left(168mn+12n+108m \right)+\sqrt{6}\left(96mn+12n+72m \right),\\
b. IN_{1}(\Upsilon)=&96mn+16n+80m+2\sqrt{30} \left(14mn+n+9m \right)+4\sqrt{6} \left(8mn+n+6m \right)\\
&+3\sqrt{2} \left(6mn+n+5m \right),\\
c. IN_{2}(\Upsilon)=&96mn+16n+80m+\frac{\sqrt{6}}{\sqrt{5}}\left(168mn+12n+108m \right)+48\sqrt{6}mn+6\sqrt{6}n+36\sqrt{6}m\\
&+36\sqrt{2}mn+6\sqrt{2}n+30\sqrt{2}m.
\end{align*}
\end{theorem}
\textbf{Proof}. Let $\Upsilon$ be the $n,m$ dimensional terpyridine complex nanosheet $TCN_{n,m} $. From Theorem 3.1, the M-polynomial of $\Upsilon$ is
\begin{align*}
M(\Upsilon; x,y)=&(96mn+16n+80m)x^2y^2+(168mn+12n+108m)x^2y^3+ (96mn+12n+72m)x^3y^3\\
   & + (36mn+6n+30m)x^3y^6.
\end{align*} 
From Table 1, we have 
\begin{align*}
& (i) \, D_{x}^{\frac{1}{2}}J(M(\Upsilon; x,y)) \\
& = D_{x}^{\frac{1}{2}}J \left[(96mn+16n+80m)x^2y^2+(168mn+12n+108m)x^2y^3 + (96mn+12n+72m)x^3y^3 \right] \\
+ & D_{x}^{\frac{1}{2}}J \left[(36mn+6n+30m)x^3y^6 \right] \\
& = D_{x}^{\frac{1}{2}} \left[(96mn+16n+80m)x^4+(168mn+12n+108m)x^5 + (96mn+12n+72m)x^6 \right] \\
& + D_{x}^{\frac{1}{2}} (36mn+6n+30m)x^9  \\
& = 2(96mn+16n+80m)x^4+\sqrt{5}(168mn+12n+108m)x^5 +\sqrt{6}(96mn+12n+72m)x^6\\
& +3(36mn+6n+30m)x^9 \\
& = (192mn+32n+160m)x^4+\sqrt{5}(168mn+12n+108m)x^5++\sqrt{6}(96mn+12n+72m)x^6 \\
& +(108mn+18n+90m)x^9. 
\end{align*}

\begin{align*}
& (ii) \, D_{x}^{\frac{1}{2}}JS_{y}^{\frac{1}{2}}S_{x}^{\frac{1}{2}}(M(\Upsilon; x,y)) \\
=&D_{x}^{\frac{1}{2}}JS_{y}^{\frac{1}{2}}S_{x}^{\frac{1}{2}} \left[(96mn+16n+80m)x^2y^2+(168mn+12n+108m)x^2y^3 \right] \\
+&D_{x}^{\frac{1}{2}}JS_{y}^{\frac{1}{2}}S_{x}^{\frac{1}{2}} \left[(96mn+12n+72m)x^3y^3 + (36mn+6n+30m)x^3y^6 \right] \\
= & D_{x}^{\frac{1}{2}}JS_{y}^{\frac{1}{2}} \left[\frac{1}{\sqrt{2}} (96mn+16n+80m)x^2y^2 + \frac{1}{\sqrt{2}}(168mn+12n+108m)x^2y^3 \right] \\
+ & D_{x}^{\frac{1}{2}}JS_{y}^{\frac{1}{2}} \left[\frac{1}{\sqrt{3}} (96mn+12n+72m)x^3y^3 + \frac{1}{\sqrt{3}}(36mn+6n+30m)x^3y^6 \right] \\
& = D_{x}^{\frac{1}{2}}J \left[\frac{1}{2} (96mn+16n+80m)x^2y^2 + \frac{1}{\sqrt{6}}(168mn+12n+108m)x^2y^3 \right] \\
& + D_{x}^{\frac{1}{2}}J \left[\frac{1}{3} (96mn+12n+72m)x^3y^3 + \frac{1}{\sqrt{18}}(36mn+6n+30m)x^3y^6 \right] \\
& = D_{x}^{\frac{1}{2}}\left[\frac{1}{2} (96mn+16n+80m)x^4+ \frac{1}{\sqrt{6}}(168mn+12n+108m)x^5 \right] \\
& + D_{x}^{\frac{1}{3}} \left[\frac{1}{2} (96mn+12n+72m)x^6 + \frac{1}{\sqrt{18}}(36mn+6n+30m)x^9 \right] \\
& = (96mn+16n+80m)x^4 + \frac{\sqrt{5}}{\sqrt{6}} (168mn+12n+108m)x^5 + \frac{\sqrt{6}}{3}(96mn+12n+72m)x^6 \\
& + \frac{1}{\sqrt{2}} (36mn+6n+30m)x^9.
\end{align*}

\begin{align*}
(iii) \, & S_{x}^{\frac{1}{2}}JD_{y}^{\frac{1}{2}}D_{x}^{\frac{1}{2}}(M(\Upsilon; x,y))  \\
&=S_{x}^{\frac{1}{2}}JD_{y}^{\frac{1}{2}}D_{x}^{\frac{1}{2}} \left[(96mn+16n+80m)x^2y^2+(168mn+12n+108m)x^2y^3 \right] \\
+ & S_{x}^{\frac{1}{2}}JD_{y}^{\frac{1}{2}}D_{x}^{\frac{1}{2}} \left[(96mn+12n+72m)x^3y^3 + (36mn+6n+30m)x^3y^6 \right] \\
&=S_{x}^{\frac{1}{2}}JD_{y}^{\frac{1}{2}} \left[\sqrt{2}(96mn+16n+80m)x^2y^2+\sqrt{2}(168mn+12n+108m)x^2y^3 \right] \\
+ & S_{x}^{\frac{1}{2}}JD_{y}^{\frac{1}{2}} \left[\sqrt{3}(96mn+12n+72m)x^3y^3 +\sqrt{3} (36mn+6n+30m)x^3y^6 \right] \\
&=S_{x}^{\frac{1}{2}}J \left[2(96mn+16n+80m)x^2y^2+\sqrt{6}(168mn+12n+108m)x^2y^3 \right] \\
+ & S_{x}^{\frac{1}{2}}J \left[3(96mn+12n+72m)x^3y^3 +\sqrt{18} (36mn+6n+30m)x^3y^6 \right] \\
&=S_{x}^{\frac{1}{2}} \left[2(96mn+16n+80m)x^4+\sqrt{6}(168mn+12n+108m)x^5 + 3(96mn+12n+72m)x^6 \right] \\
+ & S_{x}^{\frac{1}{2}} \left[\sqrt{18} (36mn+6n+30m)x^9 \right] \\
& =\frac{1}{2}(192mn+16n+80m)x^4+\frac{\sqrt{6}}{\sqrt{5}}(168mn+12n+108m)x^5 + \frac{1}{\sqrt{6}}(288mn+36n+216m)x^6 \\
+ & \frac{\sqrt{18}}{\sqrt{9}} (36mn+6n+30m)x^9.
\end{align*}

Hence, the Nirmala indices of $\Upsilon$ are given by 
\begin{align*}
(a) \, N(\Upsilon)=& D_{x}^{\frac{1}{2}}J(M(\Upsilon; x,y))|_{x=1} \\
& = 300mn+250m+50n+\sqrt{5}\left(168mn+12n+108m \right)+\sqrt{6}\left(96mn+12n+72m \right).
\end{align*} 
\begin{align*}
(b) \, IN_{1}(\Upsilon)=&D_{x}^{\frac{1}{2}}JS_{y}^{\frac{1}{2}}S_{x}^{\frac{1}{2}}(M(\Upsilon; x,y))|_{x=1}   \\
 & = 96mn+16n+80m+2\sqrt{30} \left(14mn+n+9m \right)+4\sqrt{6} \left(8mn+n+6m \right)\\
& +3\sqrt{2} \left(6mn+n+5m \right).
\end{align*}
 \begin{align*}
(c) \, IN_{2}(\Upsilon)=& S_{x}^{\frac{1}{2}}JD_{y}^{\frac{1}{2}}D_{x}^{\frac{1}{2}}(M(\Upsilon; x,y))|_{x=1}   \\
&=96mn+16n+80m+\frac{\sqrt{6}}{\sqrt{5}}\left(168mn+12n+108m \right)+48\sqrt{6}mn+6\sqrt{6}n+36\sqrt{6}m\\
&+36\sqrt{2}mn+6\sqrt{2}n+30\sqrt{2}m.
\end{align*}  

\subsection{Entropy measures of terpyridine complex nanosheet $TCN_{n,m}$}
We proceed with the computation of graph entropy metrics for the $n,m$ dimensional terpyridine complex nanosheet $TCN_{n,m}$. Initially, we assess the degree-based graph entropy expression. Subsequently, we determine the mathematical formulations for entropy measures based on Nirmala indices, utilizing the previously derived expressions of the Nirmala indices.\\\\
\textbf{Nirmala entropy of $TCN_{n,m}$:}\\
From Theorem 3.2, the  Nirmala index of $\Upsilon$ is given by
\begin{equation*}
N(\Upsilon)=300mn+250m+50n+\sqrt{5}\left(168mn+12n+108m \right)+\sqrt{6}\left(96mn+12n+72m \right).     
\end{equation*}
From Table 2 and Equation (5), we have
\begin{align*}
 ENT_{N}(\Upsilon)=&log(N(\Upsilon))  - \frac{1}{N(\Upsilon)}              \displaystyle \sum_{st \in E(\Upsilon)} \sqrt{d_{\Upsilon}(s)+d_{\Upsilon}(t)} \times log (\sqrt{d_{\Upsilon}(s)+d_{\Upsilon}(t)}) \\
 = & log(N(\Upsilon))-\frac{1}{N(\Upsilon)} \left[\displaystyle \sum_{i=1}^{4} \displaystyle \sum_{st \in E_{i}(\Upsilon)} \sqrt{d_{\Upsilon}(s)+d_{\Upsilon}(t)} \times log (\sqrt{d_{\Upsilon}(s)+d_{\Upsilon}(t)}) \right] \\
 & = log(N(\Upsilon))-\frac{1}{N(\Upsilon)} \left[(96mn+16n+80m) \cdot \sqrt{2+2} \cdot log(\sqrt{2+2}) \right] \\
 & -\frac{1}{N(\Upsilon)} \left[(168mn+12n+108m) \cdot \sqrt{2+3} \cdot log(\sqrt{2+3}) \right] \\
 - &\frac{1}{N(\Upsilon)} \left[(96mn+12n+72m) \cdot \sqrt{3+3} \cdot log(\sqrt{3+3}) \right] \\
 - &\frac{1}{N(\Upsilon)} \left[(36mn+6n+30m) \cdot \sqrt{3+6} \cdot log(\sqrt{3+6}) \right] \\
  -& \frac{1}{N(\Upsilon)}\left[(15mn-13m-13n+11) \cdot \sqrt{3+3} \cdot log(\sqrt{3+3})\right] \\
 & = log(N(\Upsilon))-\frac{1}{N(\Upsilon)} \left[(96mn+16n+80m) \cdot 2 \cdot log(2) \right] \\
 & -\frac{1}{N(\Upsilon)} \left[(168mn+12n+108m) \cdot \sqrt{5} \cdot log(\sqrt{5}) \right] \\
 - &\frac{1}{N(\Upsilon)} \left[(96mn+12n+72m) \cdot \sqrt{6} \cdot log(\sqrt{6}) + (36mn+6n+30m) \cdot 3 \cdot log(3)\right] 
  \end{align*}
Finally, we get the desired formulation of the Nirmala entropy for $TCN_{n,m}$ by substituting the value of $N(\Upsilon)$ into the previous expression.
\newpage
\textbf{First inverse Nirmala entropy of $TCN_{n,m}$:}\\
From Theorem 3.2, the first inverse Nirmala index $\Upsilon$ is given by
\begin{align*}
IN_{1}(\Upsilon)=&96mn+16n+80m+2\sqrt{30} \left(14mn+n+9m \right)+4\sqrt{6} \left(8mn+n+6m \right)\\
&+3\sqrt{2} \left(6mn+n+5m \right).    
\end{align*}
From Table 2 and Equation (6), we have
\begin{align*}
 ENT_{IN_{1}}(\Upsilon)=&log(IN_{1}(\Upsilon))  - \frac{1}{IN_{1}(\Upsilon)}              \displaystyle \sum_{st \in E(\Upsilon)} \sqrt{\frac{1}{d_{\Upsilon}(s)}+\frac{1}{d_{\Upsilon}(t)}} \times log \left( \sqrt{\frac{1}{d_{\Upsilon}(s)}+\frac{1}{d_{\Upsilon}(t)}} \right) \\
 =& log(IN_{1}(\Upsilon)) - \frac{1}{IN_{1}(\Upsilon)} \left[\displaystyle \sum_{i=1}^{4} \displaystyle \sum_{st \in E_{i}(\Upsilon)} \sqrt{\frac{1}{d_{\Upsilon}(s)}+\frac{1}{d_{\Upsilon}(t)}} \times log \left( \sqrt{\frac{1}{d_{\Upsilon}(s)}+\frac{1}{d_{\Upsilon}(t)}} \right) \right] \\
 = & log(IN_{1}(\Upsilon)) - \frac{1}{IN_{1}(\Upsilon)} \left[(96mn+16n+80m) \cdot \sqrt{\frac{1}{2}+\frac{1}{2}} \cdot log\left(\sqrt{\frac{1}{2}+\frac{1}{2}}\right) \right] \\
 - & \frac{1}{IN_{1}(\Upsilon)} \left[(168mn+12n+108m) \cdot \sqrt{\frac{1}{2}+\frac{1}{3}} \cdot log\left(\sqrt{\frac{1}{2}+\frac{1}{3}}\right) \right] \\
 - & \frac{1}{IN_{1}(\Upsilon)} \left[(96mn+12n+72m) \cdot \sqrt{\frac{1}{3}+\frac{1}{3}} \cdot log\left(\sqrt{\frac{1}{3}+\frac{1}{3}}\right) \right] \\
 - & \frac{1}{IN_{1}(\Upsilon)} \left[(36mn+6n+30m) \cdot \sqrt{\frac{1}{3}+\frac{1}{6}} \cdot log\left(\sqrt{\frac{1}{3}+\frac{1}{6}}\right) \right]
 \end{align*}
 Since $log(1)=0$, 
 \begin{align*}   
 ENT_{IN_{1}}(\Upsilon)= & log(IN_{1}(\Upsilon))  
 - \frac{1}{IN_{1}(\Upsilon)} \left[(168mn+12n+108m) \cdot \sqrt{\frac{5}{6}} \cdot log\left(\sqrt{\frac{5}{6}}\right) \right] \\
 - & \frac{1}{IN_{1}(\Upsilon)} \left[(96mn+12n+72m) \cdot \sqrt{\frac{2}{3}}  \cdot log\left(\sqrt{\frac{2}{3}}\right) \right] \\
 - & \frac{1}{IN_{1}(\Upsilon)} \left[(36mn+6n+30m) \cdot \sqrt{\frac{1}{2}} \cdot log\left(\sqrt{\frac{1}{2}}\right) \right]\\
\end{align*}
Finally, by substituting the value of $IN_{1}(\Upsilon)$ into the preceding expression, we obtain the desired formulation of the first inverse Nirmala entropy for $TCN_{n,m}$.\newpage
\textbf{Second inverse Nirmala entropy of $TCN_{n,m}$:}\\
From Theorem 3.2, the second inverse Nirmala index $\Upsilon$ is given by
\begin{align*}
IN_{2}(\Upsilon)=&96mn+16n+80m+\frac{\sqrt{6}}{\sqrt{5}}\left(168mn+12n+108m \right)+48\sqrt{6}mn+6\sqrt{6}n+36\sqrt{6}m\\
&+36\sqrt{2}mn+6\sqrt{2}n+30\sqrt{2}m.
\end{align*} 
From Table 2 and Equation (7), we have
\begin{align*}
 ENT_{IN_{2}}(\Upsilon)=& log(IN_{2}(\Upsilon))  - \frac{1}{IN_{2}(\Upsilon)}              \displaystyle \sum_{st \in E(\Upsilon)} \frac{\sqrt{d_{\Upsilon}(s) \cdot  d_{\Upsilon}(t)} }{\sqrt{d_{\Upsilon}(s) + d_{\Upsilon}(t)} } \times log \left( \frac{\sqrt{d_{\Upsilon}(s) \cdot  d_{\Upsilon}(t)} }{\sqrt{d_{\Upsilon}(s) + d_{\Upsilon}(t)} } \right) \\
 = & log(IN_{2}(\Upsilon))  - \frac{1}{IN_{2}(\Upsilon)} \left[ \displaystyle \sum_{i=1}^{4} \displaystyle \sum_{st \in E_{i}(\Upsilon)}              \frac{\sqrt{d_{\Upsilon}(s) \cdot  d_{\Upsilon}(t)} }{\sqrt{d_{\Upsilon}(s) + d_{\Upsilon}(t)} } \times log \left( \frac{\sqrt{d_{\Upsilon}(s) \cdot  d_{\Upsilon}(t)} }{\sqrt{d_{\Upsilon}(s) + d_{\Upsilon}(t)} } \right) \right] \\
 = & log(IN_{2}(\Upsilon))  - \frac{1}{IN_{2}(\Upsilon)} \left[ (96mn+16n+80m) \cdot \frac{\sqrt{4}}{\sqrt{4}} \cdot log  \left(\frac{\sqrt{4}}{\sqrt{4}} \right) \right]\\
  & - \frac{1}{IN_{2}(\Upsilon)} \left[ (168mn+12n+108m) \cdot \frac{\sqrt{6}}{\sqrt{5}} \cdot log  \left(\frac{\sqrt{6}}{\sqrt{5}} \right) \right]\\
 - & \frac{1}{IN_{2}(\Upsilon)} \left[(96mn+12n+72m) \cdot \frac{\sqrt{9}}{\sqrt{6}} \cdot log \left(\frac{\sqrt{9}}{\sqrt{6}} \right) \right] \\
 - & \frac{1}{IN_{2}(\Upsilon)} \left[(36mn+6n+30m) \cdot \frac{\sqrt{18}}{\sqrt{9}} \cdot log \left(\frac{\sqrt{18}}{\sqrt{9}} \right) \right]
 \end{align*} 
Since $log(1)=0$, 
\begin{align*}
ENT_{IN_{2}}(\Upsilon)=&log(IN_{2}(\Upsilon))  - 
  \frac{1}{IN_{2}(\Upsilon)} \left[ (168mn+12n+108m) \cdot \frac{\sqrt{6}}{\sqrt{5}} \cdot log  \left(\frac{\sqrt{6}}{\sqrt{5}} \right) \right]\\
 - & \frac{1}{IN_{2}(\Upsilon)} \left[(96mn+12n+72m) \cdot \frac{\sqrt{9}}{\sqrt{6}} \cdot log \left(\frac{\sqrt{9}}{\sqrt{6}} \right) \right] \\
 - & \frac{1}{IN_{2}(\Upsilon)} \left[(36mn+6n+30m) \cdot \frac{\sqrt{18}}{\sqrt{9}} \cdot log \left(\frac{\sqrt{18}}{\sqrt{9}} \right) \right]
\end{align*}
The second inverse Nirmala entropy for $TCN_{n,m}$ can finally be expressed as desired by substituting the value of $IN_{2}(\Upsilon)$ in the previous expression.
 \section{Comparison through numerical and graphical demonstrations}
 Graph entropy metrics are widely used in many scientific domains, including computer science, information theory, chemistry, biological therapies, and pharmacology. Therefore, scientists working in these disciplines rely on numerical calculation and graphical representation to appropriately characterize these molecular characteristics. This section compares the Nirmala indices and the corresponding entropy measures using numerical computation and 2D line graphs. 
 
 Table 3 shows the results of the numerical calculation of the Nirmala indices and corresponding entropy measures for $TCN_{n,m}$. Table 3 includes the intervals $1\leq n \leq 25$ and $1 \leq m \leq 25$, assuming $n=m$. Additionally, Figure 3 uses 2D line graphs for $1\leq n \leq 25$ and $1 \leq m \leq 25$ to compare the Nirmala indices and the corresponding entropy metrics. 
 
 From Table 3 and Figure 3, the following two remarks are possible. \\
\textbf{Remark 1:}  The Nirmala indices and associated entropy measures of the terpyridine complex nanosheet $TCN_{n,m}$ increase as the values of $n$
and $m$ increase. \\
\textbf{Remark 2:} For terpyridine complex nanosheet $\Upsilon=TCN_{n,m}$, we have the following inequality relationships: \\
$IN_1(\Upsilon) < IN_2(\Upsilon) < N(\Upsilon)$; and  
$ENT_{N}(\Upsilon) \approx ENT_{IN_{1}}(\Upsilon) \approx ENT_{IN_{2}}(\Upsilon)$.
 \newpage
\textbf{Table 3}. Calculated values of the Nirmala indices and their associated entropy measures of $TCN_{n,m}$, where $1 \leq n \leq 25, 1 \leq m \leq 25$ with $n=m$. 

 \begin{center}
\begin{table}[h!]
\centering
\renewcommand{\arraystretch}{1.5}
\begin{tabular}{|>{\centering\arraybackslash}m{2cm}||>{\centering\arraybackslash}m{2cm} |>{\centering\arraybackslash}m{2cm} |>{\centering\arraybackslash}m{2cm} |>{\centering\arraybackslash}m{2cm} |>{\centering\arraybackslash}m{2cm} |>{\centering\arraybackslash}m{2cm} |>{\centering\arraybackslash}m{1.8cm} ||} 
 \hline
 [n,m] & N & $IN_{1}$ & $IN_{2}$ & $ENT_{N}$ &  $ENT_{IN_{1}}$  &   $ENT_{IN_{2}}$
 \\ [0.5ex] 
 \hline\hline
[1,1] & 1684.89 & 652.78 & 829.76 & 6.5886 & 6.5906 & 6.5901 \\
\hline
[2,2] & 5191.41 & 2011.97 & 2556.57 & 7.7144 & 7.7163 & 7.7158  \\ 
\hline
[3,3] & 10519.54 & 4077.57 & 5180.42 & 8.4208 & 8.4227 & 8.4222 \\ 
 \hline 
[4,4] & 17669.30 & 6849.57 & 8701.32 & 8.9395 & 8.9414 & 8.9409 \\ 
\hline 
[5,5] & 26640.68 & 10327.97 & 13119.26 & 9.3502 & 9.3521 & 9.3516 \\
\hline 
[6,6] & 37433.68 & 14512.78 & 18434.25 & 9.6904 & 9.6923 & 9.6918 \\
\hline
[7,7] & 50048.30 & 19403.99 & 24646.28 & 9.9809 & 9.9827 & 9.9822 \\
\hline
[8,8] & 64484.55 & 25001.60 & 31755.35 & 10.2343 & 10.2362 & 10.2357 \\
\hline 
[9,9] & 80742.41 & 31305.62 & 39761.47 & 10.4592 & 10.4611 & 10.4605\\
\hline 
[10,10] & 98821.89 & 38316.04 & 48664.63 & 10.6613 & 10.6631 & 10.6626 \\
\hline
[11,11] & 118723.00 & 46032.86 & 58464.83 & 10.8448 & 10.8466 & 10.8461 \\
\hline 
[12,12] & 140445.72 & 54456.09 & 69162.08 & 11.0128 & 11.0147 & 11.0142 \\
\hline 
[13,13] & 163990.07 & 63585.72 & 80756.38 & 11.1678 & 11.1697 & 11.1691 \\
\hline 
[14,14] & 189326.03 & 73421.76 & 93247.71 & 11.3116 & 11.3135 & 11.3130 \\
\hline
[15,15] & 216543.62 & 83964.20 & 106636.09 & 11.4458 & 11.4477 & 11.4472 \\
\hline
[16,16] & 245552.83 & 95213.04 & 120921.52 & 11.5715 & 11.5734 & 11.5729 \\
\hline 
[17,17] & 276383.66 & 107168.29 & 136103.99 & 11.6898 & 11.6917 & 11.6912 \\
\hline 
[18,18] & 309036.16 & 119829.94 & 152183.50 & 11.8015 & 11.8034 & 11.8028 \\
\hline 
[19,19] & 343510.18 & 133197.99 & 169160.06 & 11.9073 & 11.9091 & 11.9086 \\
\hline 
[20,20] & 379805.88 & 147272.45 & 187033.66 & 12.0077 & 12.0096 & 12.0091 \\
\hline
[21,21] & 417923.19 & 162053.31 & 205804.43 & 12.1034 & 12.1052 & 12.1047\\
\hline
[22,22] & 457862.12 & 177540.58 & 225471.77 & 12.1946 & 12.1965 & 12.1960 \\
\hline
[23,23] & 499622.26 & 193734.24 & 246036.72 & 12.2819 & 12.2838 & 12.2833  \\
\hline
[24,24] & 543204.85 & 210634.32 & 267498.50 & 12.3656 & 12.3674 & 12.3669 \\
\hline
[25,25] & 588608.65 & 228240.79 & 289857.32 & 12.4458 & 12.4477 & 12.4472 \\
\hline
\end{tabular}
\end{table} 
\end{center}
\newpage
\begin{figure}[h!]
\centering
\includegraphics[width=150mm]{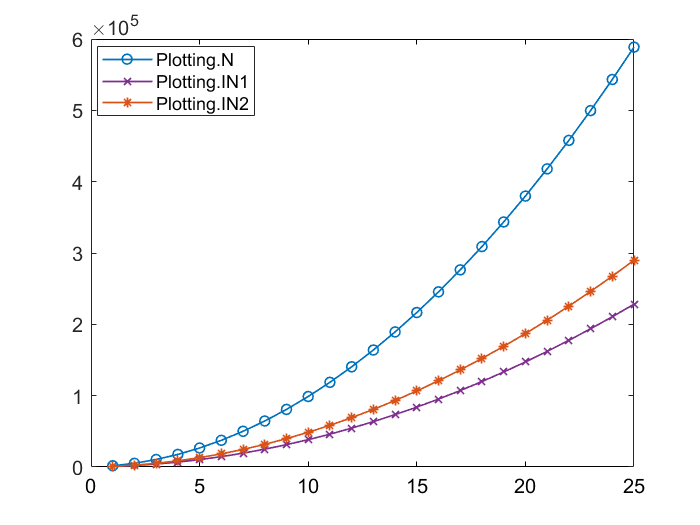}  
  \end{figure} 
  \begin{figure}[h!]
\centering
\includegraphics[width=145mm]{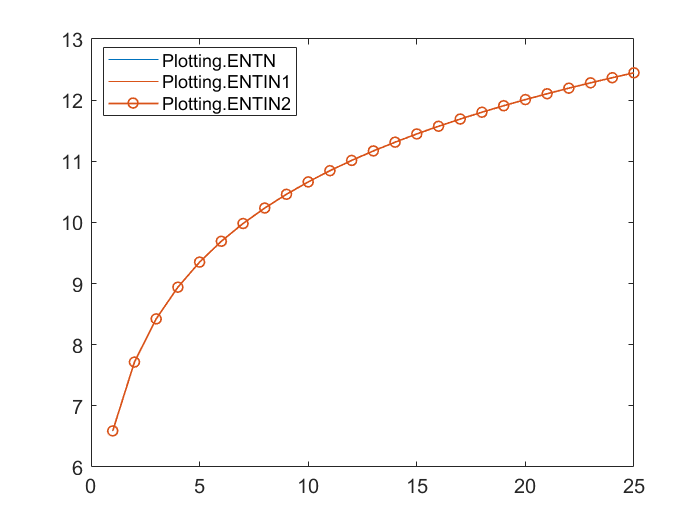}  
  \end{figure} 
\textbf{Figure 3}. Comparison of the Nirmala indices and their associated entropy measures of $TCN_{n,m}$ through a 2D line plot for $1 \leq n \leq 25, 1 \leq m \leq 25$ with $n=m$.
\newpage
\section{The logarithmic regression model}
We use logarithmic regression analysis to examine the connection between the dependent variable and one or more predictor variables in our dataset. Logarithmic regression is a nonlinear regression technique that modifies the dependent or predictor variables by using logarithmic functions. The logarithmic regression model equation is as follows.
\begin{equation*}
y=a*log(x)+b, 
\end{equation*} 
where the response variable is $y$ and the predictor variable is $x$. The relationship between $x$ and $y$ is represented by the regression coefficients $a$ and $b$.

Here, using a logarithmic regression analysis, we investigate the relationship between the Nirmala indices and entropy metrics of the terpyridine complex nanosheet $TCN_{n,m}$ for $1 \leq n, m \leq 25$, with $m=n$. The statistical measurements employed in the study include the squared correlation coefficient ($R^{2}$), root mean square error (RMSE), adjusted squared correlation coefficient (Adj. R-sq), and sum of square error (SSE). While a higher $R^{2}$ value (near 1) suggests that the regression line fits the data better, a low RMSE value (closer to 0) indicates that the model operates effectively. Getting a lower RMSE value is our primary objective in this case. 

The statistics of curve fitting of the Nirmala indices versus Nirmala entropy measures for terpyridine complex nanosheet $TCN_{n,m}$ using the logarithmic regression are shown in Table 4.
\vspace{5mm}
\begin{table}[h!]
\centering
\renewcommand{\arraystretch}{2.5}
\begin{tabular}{|>{\centering\arraybackslash}m{7cm}||>{\centering\arraybackslash}m{1cm} |>{\centering\arraybackslash}m{2.3cm} |>{\centering\arraybackslash}m{1cm} |>{\centering\arraybackslash}m{3cm} ||} 
 
 \hline
 Model & $R^{2}$ & SSE & Adj. R-sq & RMSE
 \\ [0.5ex] 
 \hline\hline
 $ENT_{N}=1.0002*log(N)-0.8417$  &  1   & 0.000000248 & 1 & \textbf{0.00010389} \\
\hline
 $ENT_{IN_{1}}=1.00*log(IN_{1})+0.1092$  & 1   & 0.0000000144 & 1 & \textbf{0.000025025} \\
\hline
 $ENT_{IN_{2}}=1.0002*log(IN_{2})-0.1318$  & 1   & 0.000000162 & 1 & \textbf{0.000084135}  \\
 \hline
\end{tabular}
\end{table} 
  
\textbf{Table 4}. Statistics of curve fitting of the Nirmala indices vs. Nirmala entropy measures of terpyridine complex nanosheet $TCN_{n,m}$.
\newpage
\begin{figure}[h!]
\centering
\includegraphics[width=140mm]{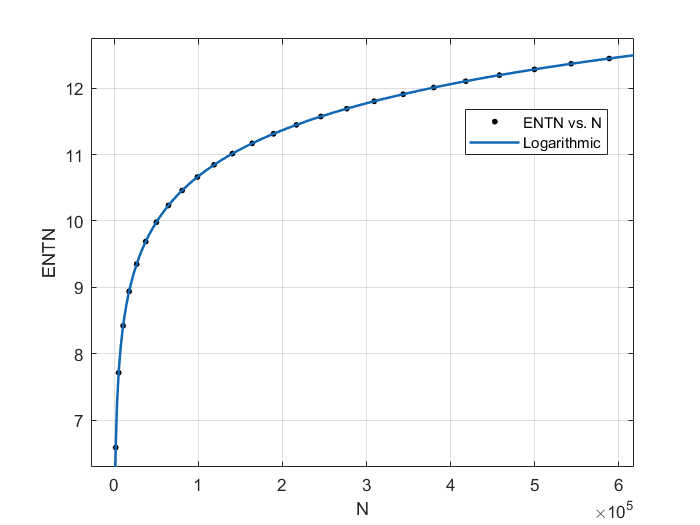}  
  \end{figure} 
  \begin{figure}[h!]
\centering
\includegraphics[width=140mm]{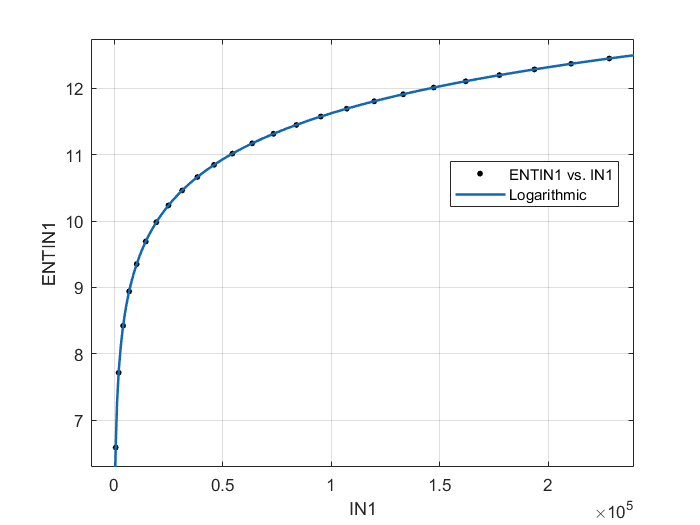}  
  \end{figure} 
  \newpage
 \begin{figure}[h!]
\centering
\includegraphics[width=140mm]{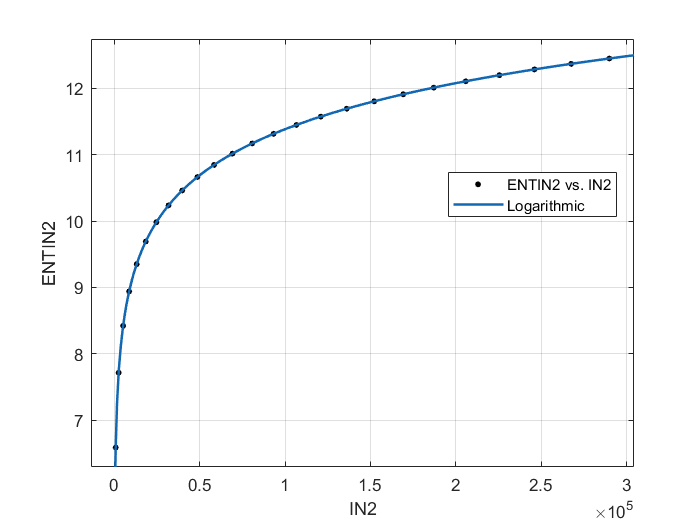}  
  \end{figure} 
  \textbf{Figure 4}. Curve fitting plots for the Nirmala indices vs. Nirmala entropy measures of terpyridine complex nanosheet $TCN_{n,m}$ for 
  $1 \leq n, m \leq 25$ with $m=n$.
 
 \section{Discussion}
 
Topological indices, which are numerical representations of molecular groups in graph theory, help researchers better understand the molecular properties and behavior of chemical and biological systems. Precise computation of numerical indices provides researchers with useful data that improves their understanding of the subject. 

In this work, we investigate the so-called Nirmala indices of the terpyridine complex nanosheet $TCN_{n,m}$, which are degree-based topological indices. Table 3 and Figure 3 demonstrate how the Nirmala indices and corresponding entropy measures of $TCN_{n,m}$ increase as $n$ and $m$ increase. Entropy measures help determine the complexity and distribution of a dataset by evaluating its uncertainty or information content. They are widely used in information theory, thermodynamics, and data analysis. Accurate computations of numerical entropy provide researchers with important information that enhances their comprehension of the network under study. This justifies our focus on the edge weight entropy of $TCN_{n,m}$, considering its advantages. The entropy measurements of the terpyridine complex nanosheet $TCN_{n,m}$ increase as $n$ and $m$ grow, as Table 3 and Figure 3 demonstrate. 

\newpage
 
Logarithmic regression is a non-linear regression technique used in the social sciences, biology, and economics to account for non-linearity in data and provide predictions. Table 4 shows the statistical parameters used in the study, including the squared correlation coefficient ($R^{2}$), sum of square error (SSE), adjusted squared correlation coefficient (Adj. R-sq), and root mean square error (RMSE). A higher $R^{2}$ value (closer to 1) indicates that the regression line fits the data better. From Figure 4, we can see that the Nirmala indices and the corresponding entropy measure values of $TCN_{n,m}$ look good and fit over the curve as well.

\section{Conclusion}
This research work has made use of the definitions of the Nirmala indices and entropy measures derived from them. A mathematical formulation of the Nirmala indices of the terpyridine complex nanosheet $TCN_{n,m}$ has been achieved. The entropy measures of $TCN_{n,m}$ based on the Nirmala indices have been analyzed using its M-polynomial. Furthermore, we have compared the Nirmala indices and associated entropy measures using 2D line plots, which have been numerically generated and displayed. Table 3 and Figure 3 show that when $n$ and $m$ increase, the Nirmala indices and related entropy measurements of the terpyridine complex nanosheet $TCN_{n,m}$ also increase. Additionally, the inequality relationships shown below are also accurate. 
\begin{center}
$IN_1(\Upsilon) < IN_2(\Upsilon) < N(\Upsilon)$; and  
$ENT_{N}(\Upsilon) \approx ENT_{IN_{1}}(\Upsilon) \approx ENT_{IN_{2}}(\Upsilon)$.
\end{center}

The topology and structural characteristics of the terpyridine complex nanosheet $TCN_{n,m}$ will be examined in the fields of electronic, mechanical, optical, and nanoelectronic technology with the help of the study's results.

\section*{Funding} No funding is available for this study.
\section*{Data Availability Statement}
This manuscript has no associated data.
\section*{Declarations}
\textbf{Conflict of interest} The author declares that he has no known competing financial interests or personal relationships that could have appeared to influence the work reported in this paper.

  \makeatletter
\renewcommand{\@biblabel}[1]{[#1]\hfill}

\makeatother

\end{document}